\documentclass[twoside]{mhd}
\usepackage[latin9]{inputenc}
\usepackage{array}
\usepackage{float}
\usepackage{units}
\usepackage{multirow}
\usepackage{amstext}
\usepackage{graphicx}
\usepackage{esint}

\makeatletter

\providecommand{\tabularnewline}{\\}

\mhdhead{53}{1}{1}

\makeatother

\begin{document}
\title{A homopolar disc dynamo experiment\\ with liquid metal contacts}
\author{R. A. Avalos-Zúñiga\inst{1}, J. Priede\inst{2},  C. E. Bello-Morales\inst{1}}
\institute{ CICATA-Qro, Instituto Politécnico Nacional, Cerro Blanco 141, Colinas del Cimatario, 76090, Querétaro, Mexico \and Applied Mathematics Research Centre, Coventry University, Coventry, CV1 5FB, UK} 

\maketitle
\begin{abstract}
We present experimental results of a homopolar disc dynamo constructed
at CICATA-Querétaro in Mexico. The device consists of a flat, multi-arm
spiral coil which is placed above a fast-spinning metal disc and connected
to the latter by sliding liquid-metal electrical contacts. Theoretically,
self-excitation of the magnetic field is expected at the critical
magnetic Reynolds number $Rm\approx45,$ which corresponds to a critical
rotation rate of about $\unit[10]{Hz.}$ We measured the magnetic
field above the disc and the voltage drop on the coil for the rotation
rate up to $\unit[14]{Hz,}$ at which the liquid metal started to
leak from the outer sliding contact. Instead of the steady magnetic
field predicted by the theory we detected a strongly fluctuating magnetic
field with a strength comparable to that of Earth's magnetic field
which was accompanied by similar voltage fluctuations in the coil.
These fluctuations seem to be caused by the intermittent electrical
contact through the liquid metal. The experimental results suggest
that the dynamo with the actual electrical resistance of liquid metal
contacts could be excited at the rotation rate of around $\unit[21]{Hz}$
provided that the leakage of liquid metal is prevented. 
\end{abstract}

\section*{Introduction}

The homopolar disc dynamo is one of the simplest models of the self-excitation
of magnetic field by moving conductors which is often used to illustrate
the dynamo effect \cite{MoffatH.K:Cambrige:1978,BechBranderburg:AnnuRevAstronAstrphys:34}.
In its simplest form originally considered by Bullard \cite{Bullard:PhilSoc:51},
the dynamo consists of a solid metal disc which rotates about its
axis and a wire twisted around it which is connected through sliding
contacts to the rim and to the axis of the disc. For a sufficiently
high rotation rate, the voltage induced by the rotation of the disc
in a magnetic field generated by an initial current perturbation can
exceed the voltage drop due to the ohmic resistance. At this point,
the initial perturbation starts to grow exponentially leading to the
self- excitation of current and the associated magnetic field. Despite
its simplicity, no successful experimental implementation of the disc
dynamo is known so far. The problem appears to be the sliding electrical
contacts which are required to convey the current between the rim
and the axis of the rotating disc. The electrical resistance of the
sliding contacts, usually made of solid graphite brushes, is typically
by several orders of magnitude higher than that of the rest of the
setup \cite{RaedlerRheinhardt:MHD:38}. This results in unrealistically
high rotation rates which are required for dynamo to operate. To overcome
this problem \textit{Priede et al.} \cite{Priede=000026Avalos:PhysLetA:377}
proposed a theoretical design of homopolar disc dynamo which uses
liquid-metal sliding electrical contacts similar to those employed
in homopolar motors and generators \cite{MariboGavrilashReillyLynchSondergaard:1-7:2010,CebersMaiorov:MaGyd:80:3}.
The design consists of a flat multi-arm spiral coil placed above a
fast-spinning copper disc and connected to the latter by sliding liquid-metal
electrical contacts. The theoretical results obtained in \cite{Priede=000026Avalos:PhysLetA:377}
show a minimum magnetic Reynolds number of $40$, at which the dynamo
is self-excited. This can be achieved by using a copper disc of $\unit[60]{cm}$
in diameter which rotates with the frequency of $\unit[10]{Hz.}$
In this paper, we describe a laboratory model such a disc dynamo and
report the first results of its operation.

\section{Description of experimental setup}

\begin{figure}[H]
\noindent \centering{}\includegraphics[bb=-200bp 0bp 705bp 550bp,clip,width=0.5\columnwidth]{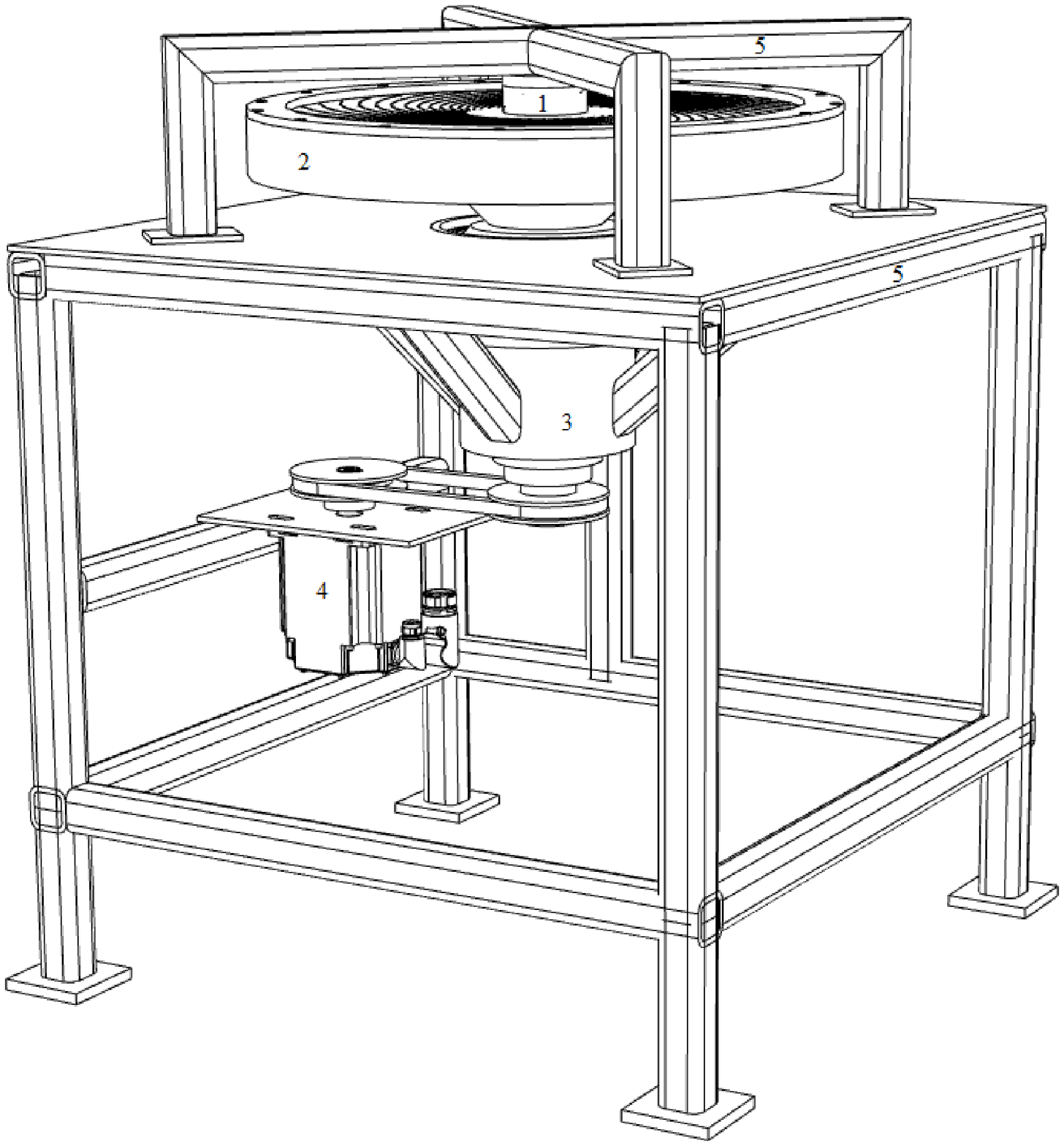}\put(-140,0){(a)}\includegraphics[bb=0bp 110bp 720bp 420bp,clip,width=0.5\columnwidth]{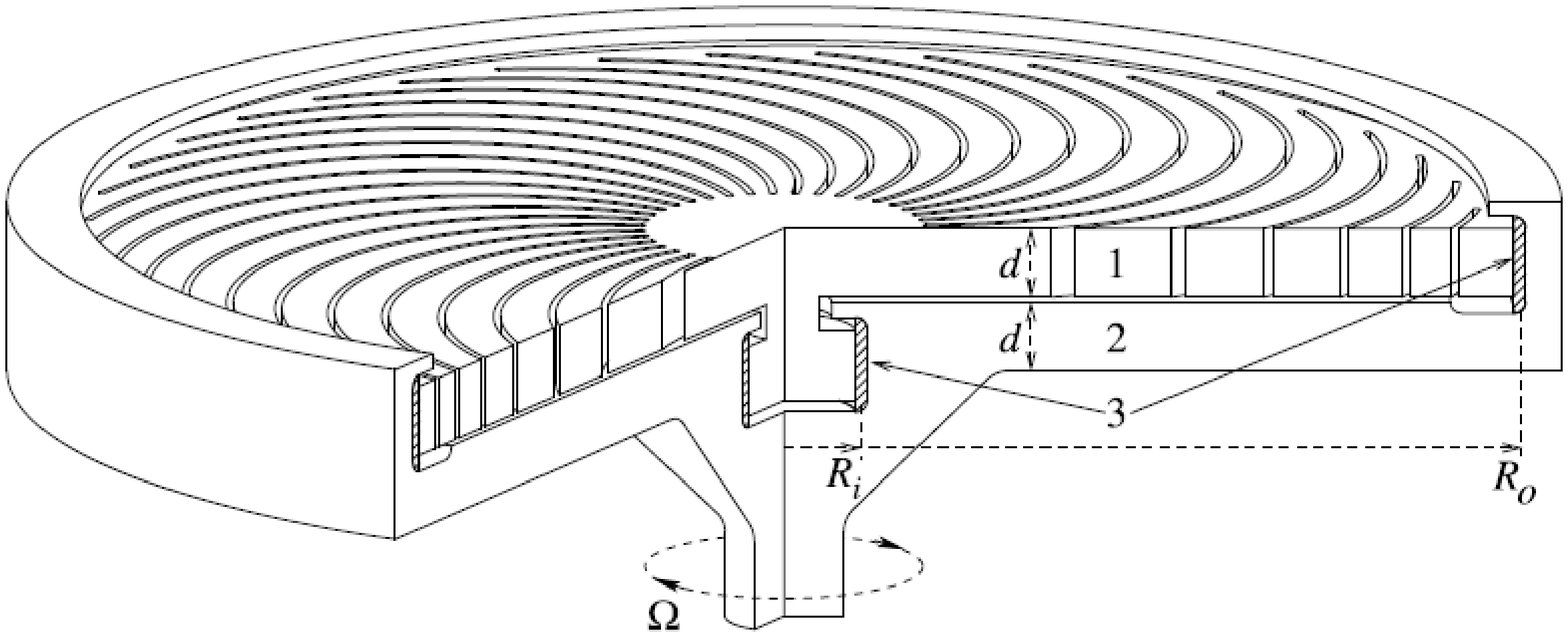}\put(-170,0){(b)}\caption{\label{fig:schm}(a) Schematic view of the dynamo facility showing
stationary coil (1), fast-spinning disc (2), gearing system (3), AC
motor (4), and iron frame(5); (b) cross-sectional view of the disc
dynamo setup consisting of a stationary coil (1) made of a copper
disc sectioned by spiral slits and a fast-spinning disc (2), which
is electrically connected to the former by sliding liquid-metal contacts
(hatched) (3).}
\end{figure}
The dynamo setup shown in Fig. \ref{fig:schm} consists of a fast-spinning
copper disc and a coil made of a stationary copper disc which is sectioned
by $40$ logarithmic spiral slits with a constant pitch angle $\alpha\approx58^{\circ}$
relative to the radial direction. The rotating disc includes a central
cylindrical cavity and an annular cavity at its rim. The coil disc
has a cylindrical solid electrode protruding $\unit[4]{cm}$ out from
the center of its bottom face. It is inserted $\unit[3]{mm}$ over
the rotating disc surface forming the inner and outer annular gaps
of width $\unit[\delta=3]{mm}$ and height $\unit[d=3]{cm}$, with
inner and outer radii $R_{i}=\unit[4.5]{cm}$  and $\unit[R_{o}=30]{cm,}$
respectively. At the bottom of the annular gaps, there is an eutectic
alloy of $GaInSn$ which is liquid at the room temperature. As the
rotation rate of the disc increases the centrifugal force pushes the
liquid metal through the gaps and creates sliding contacts which connect
electrically the stationary coil and the rotating disc. At the top
of the sliding contact, a ring cap with $40$ thin vanes at its bottom
face is attached to the rotating disc. This creates a centrifugal
pump which acts as a dynamic seal for the liquid metal. The coil and
disc are hold by iron supports which are electrically isolated from
the copper discs. The spinning disc is driven by a $\unit[3]{hp}$
AC motor through a gearing system. The motor rotation rate is controlled
by a variable frequency drive (VFD) from $0.5$ to $\unit[20]{Hz}$
which correspond to $Rm=2\cdots86.$

\section{Magnetic field and voltage measurements}

\begin{figure}[H]
\begin{centering}
\includegraphics[width=0.6\columnwidth]{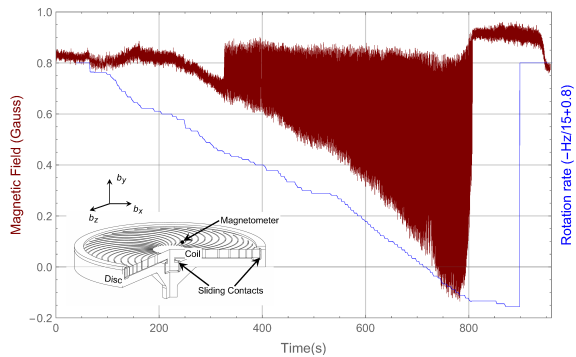}
\par\end{centering}
\caption{\label{fig:frqB-t}Rotation rate and the vertical component ($B_{y}$)
of the magnetic field versus time.}
\end{figure}
\begin{figure}[H]
\begin{centering}
\includegraphics[width=0.6\columnwidth]{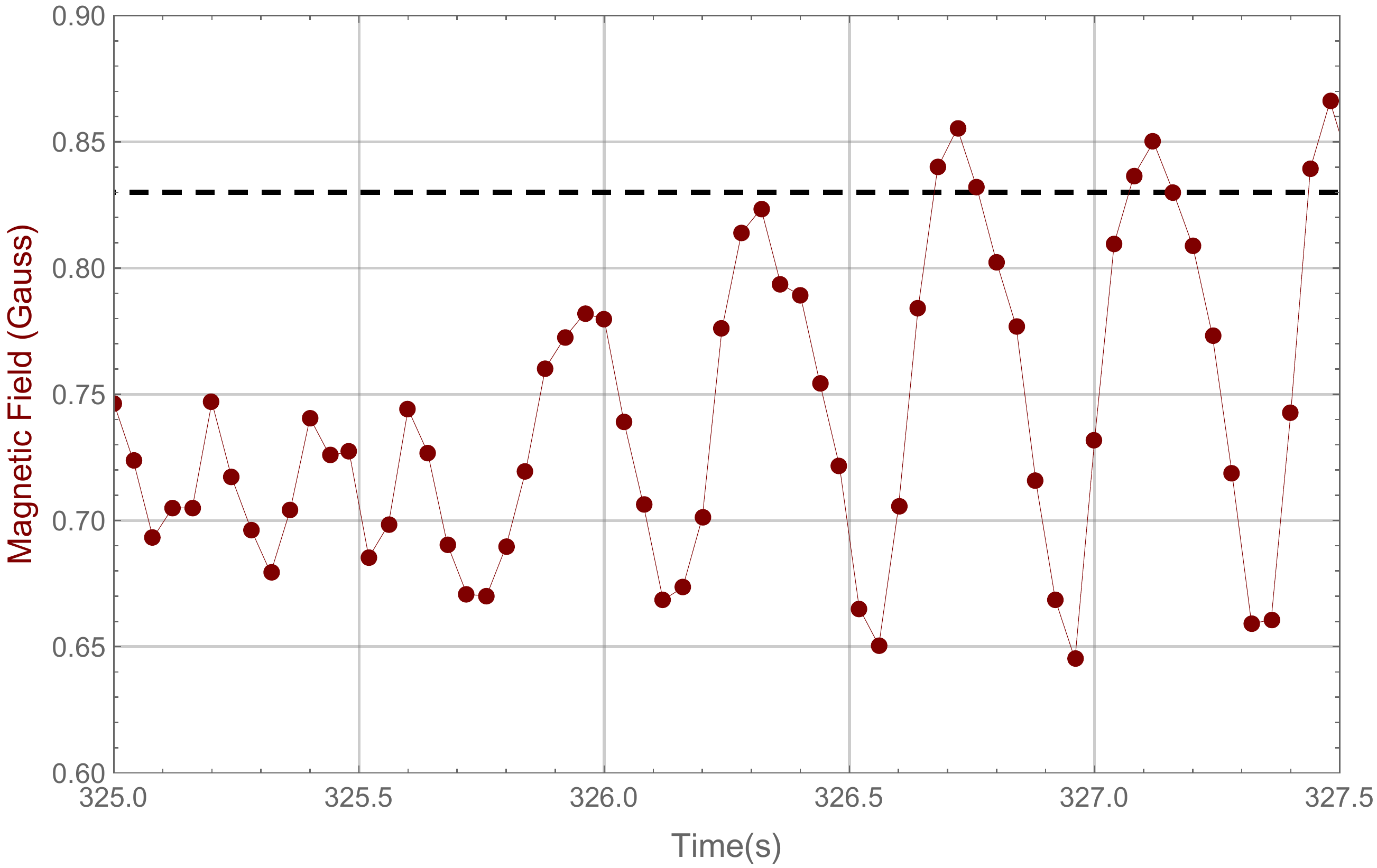}
\par\end{centering}
\caption{\label{fig:b-t}Temporal variation of the vertical component ($b_{y}$)
of the magnetic field for the rotation rate of $\unit[5.45]{Hz}$.
Dashed line corresponds to the background magnetic field.}
\end{figure}

We report measurements of the temporal variation of the magnetic field
and the induced voltage at the upper face of the coil for rotation
rates of the disc ranging from $0.5$ to $\unit[14]{Hz.}$ The sense
of rotation is opposite to the orientation of the spiral arms of the
coil which corresponds to a clock-wise rotation in the setup shown
in Fig. \ref{fig:schm}(b). The measurements of the induced magnetic
field were performed with a three-axis Hall sensor (Low-Field METROLAB
magnetometer 1176LF) placed at the upper face of the coil close to
the inner radius of the spiral slits. The rate of data acquisition
was $30$ samples per second for a period of $14$ minutes. In Figure
\ref{fig:frqB-t} we show the temporal variation of the vertical component
($b_{y}$) of the measured magnetic field together with the rotation
rate of the disc. For a rotation frequency of $\unit[4.5]{Hz}$ the
field starts to decrease nearly linearly with the increase of the
rotation rate. Then around the rotation frequency of $\unit[5.45]{Hz}$
the field starts to oscillate between $\unit[0.66-0.87]{G}$ and $\unit[0.735-0.75]{G}$
in approximately one second (see Fig. \ref{fig:b-t}). During the
maximum phase, the field returns close its background value, which
was measured at low rotation rates. The background magnetic field,
which is important for the interpretation of experimental results,
will be discussed in the next section. Such oscillations persist up
to the rotation frequency of $\unit[13.5]{Hz}$. At this point the
period of field oscillations abruptly increased to $22.5$ seconds
and stayed nearly the same as the disc rotation rate was continuously
increased up to its highest value of $\unit[14]{Hz.}$

\begin{figure}[H]
\begin{centering}
\includegraphics[width=0.6\columnwidth]{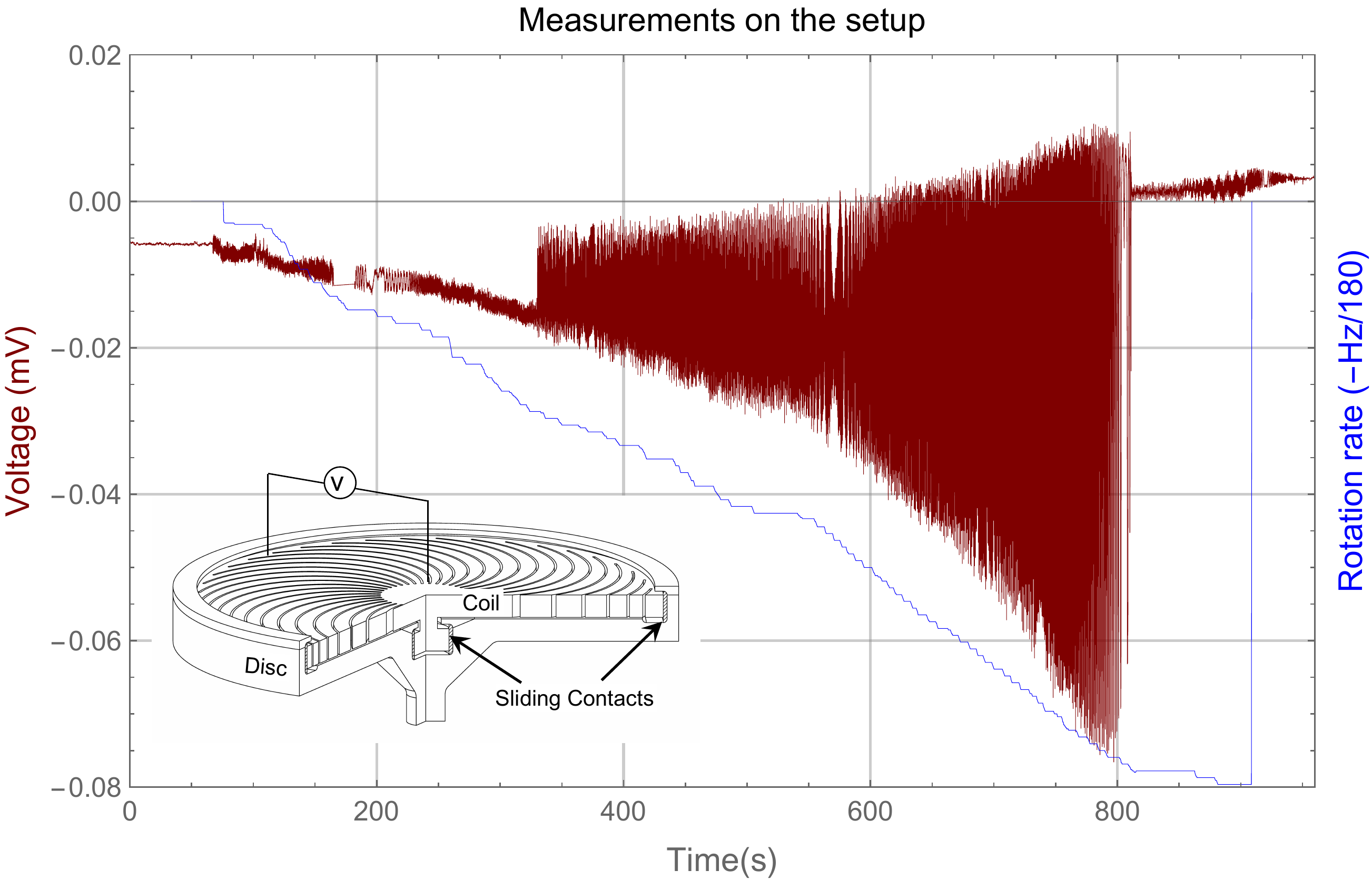}
\par\end{centering}
\caption{\label{fig:frqvlt-t}Rotation rate and the coil voltage versus time.}
\end{figure}

Voltage was measured at the upper face of the coil between the inner
and outer radius of spiral arm (see sketch in Fig. \ref{fig:frqvlt-t})
using a digital multimeter (Keithley 2100). The data was acquired
with a rate of $10$ samples per second for a period of $14$ minutes.
In Figure \ref{fig:frqvlt-t}, we show the temporal variation of the
induced voltage together with the rotation rate of the disc. Contrary
to the magnetic field, the variation of the induced voltage was observable
as soon as the disc started to rotate. At the rotation frequencies
higher than $\unit[5]{Hz}$ the voltage was found to vary similarly
to the magnetic field.

\section{Interpretation of experimental data}

In order to interpret the experimental results presented above we
need take into account that they were obtained in the presence of
a rather complicated background magnetic field. First of all, there
was Earth\textquoteright s magnetic field with a downward (negative)
vertical component of strength $\unit[0.32]{G.}$ However, the Hall
sensor, which was attached close to the iron frame holding the coil,
measured a background magnetic field with an upward (positive) vertical
component of strength $\unit[0.84]{G.}$ In the absence of more detailed
measurements it seems reasonable to assume that the measured background
magnetic field is mostly due to the magnetization of the iron frame
and, thus, localized in its vicinity while the disc is largely subject
to Earth\textquoteright s magnetic field. Then a subcritical amplification
of Earth\textquoteright s magnetic field, which is opposite to that
of iron frame at the sensor location, may explain the observed reduction
in the measured magnetic field. It has to be noted that there is only
amplification but no flux expulsion effect in our dynamo which could
provide an alternative explanation. In order to test this hypothesis,
let us estimate the current, voltage and magnetic field induced by
the disc rotating with a subcritical (below the dynamo threshold)
frequency in Earth\textquoteright s magnetic field. Taking into account
the background magnetic field with the vertical component $B_{0}$
and the associated angular flux density $\phi_{0}=\int_{R_{i}}^{R_{o}}B_{0}r\,dr\approx B_{0}\left(R_{o}^{2}-R_{i}^{2}\right)/2$,
the voltage induced across the disc rotating with angular velocity
$\Omega$ can be written according to the theoretical model of\textit{
Priede et al.} \cite{Priede=000026Avalos:PhysLetA:377} as

\[
\Delta\varphi_{d}=\Omega\left(\phi_{0}+\phi_{d}\right)+\frac{I_{0}\ln(\lambda)}{2\pi\sigma d},
\]
where 
\[
\phi_{d}=\frac{\beta\text{\ensuremath{\mu_{0}}}I_{0}R_{o}}{8\pi^{2}}\bar{\phi}(\lambda)
\]
 is the magnetic flux density induced by the current $I_{0}$ flowing
across the copper coil with the conductivity $\sigma=\unit[5.96\times10^{7}]{S/m}$
and the helicity $\beta=\tan(58^{\circ})\approx1.6$ of the logarithmic
spiral arms; $\lambda=R_{0}/R_{i}=1/6$ is the radii ratio for which
$\bar{\phi}(\lambda)$   the dimensionless flux $\bar{\phi}(\lambda)\approx1.7$
\cite{Priede=000026Avalos:PhysLetA:377}; $\mu_{0}=\unit[4\pi\times10^{-7}]{H/m}$
is the permeability of vacuum.  The current is related with the coil
voltage by Ohm's law: 
\[
\Delta\varphi_{c}=-I_{0}\frac{(1+\beta^{2})\ln\lambda}{2\pi\sigma d}.
\]
which allows us to determine $I_{0}$ and other related quantities
from the measured values of $\Delta\varphi_{c}.$ On the other hand,
applying Ohm\textquoteright s law to the whole electrical circuit
including the disc, coil and liquid-metal contacts we obtain 
\[
\Delta\varphi_{d}-\Delta\varphi_{c}=I_{0}\mathcal{R},
\]
where $\mathcal{R}$ is a liquid-metal contact resistance. This relation
allows us to evaluate the actual contact resistance from the measured
coil voltage. The results are shown in Table \ref{tab:omegc} in terms
of the dimensionless contact resistance $\bar{\mathcal{R}}=2\pi\sigma d\,\mathcal{R}$,
which is seen to be about an order of magnitude higher than its theoretical
estimate $\bar{\mathcal{R}}\approx1.26$ \cite{Priede=000026Avalos:PhysLetA:377}.
Such a high contact resistance may due to the heavy oxidation of the
eutectic alloy of GaInSn which was observed in experiments. As a result,
the rotation rates at which dynamo could become self-sustained are
around $\unit[25]{Hz}$ which are about a factor of two higher than
the theoretical value.
\begin{table}[H]
\begin{centering}
\begin{tabular}{|c|c|c|c|}
\hline 
\multirow{1}{*}{$\Omega$(Hz)} & $\bar{\mathcal{R}}$ & $Rm_{c}$ & $\Omega_{c}$(Hz)\tabularnewline
\hline 
\hline 
6 & 12.3 & 97 & 22.9\tabularnewline
\hline 
10 & 15.1 & 110 & 25.9\tabularnewline
\hline 
14 & 15.2 & 110.5 & 26.1\tabularnewline
\hline 
\end{tabular}
\par\end{centering}
\caption{\label{tab:omegc}The dimensionless contact resistance $\bar{\mathcal{R}}$
and the corresponding critical values $Rm_{c}$ and $\Omega_{c}$
evaluated using the coil voltage measured at different rotation rates.}
\end{table}

\begin{figure}[H]
\centering{}\includegraphics[bb=0bp 0bp 760bp 485bp,clip,width=0.5\columnwidth]{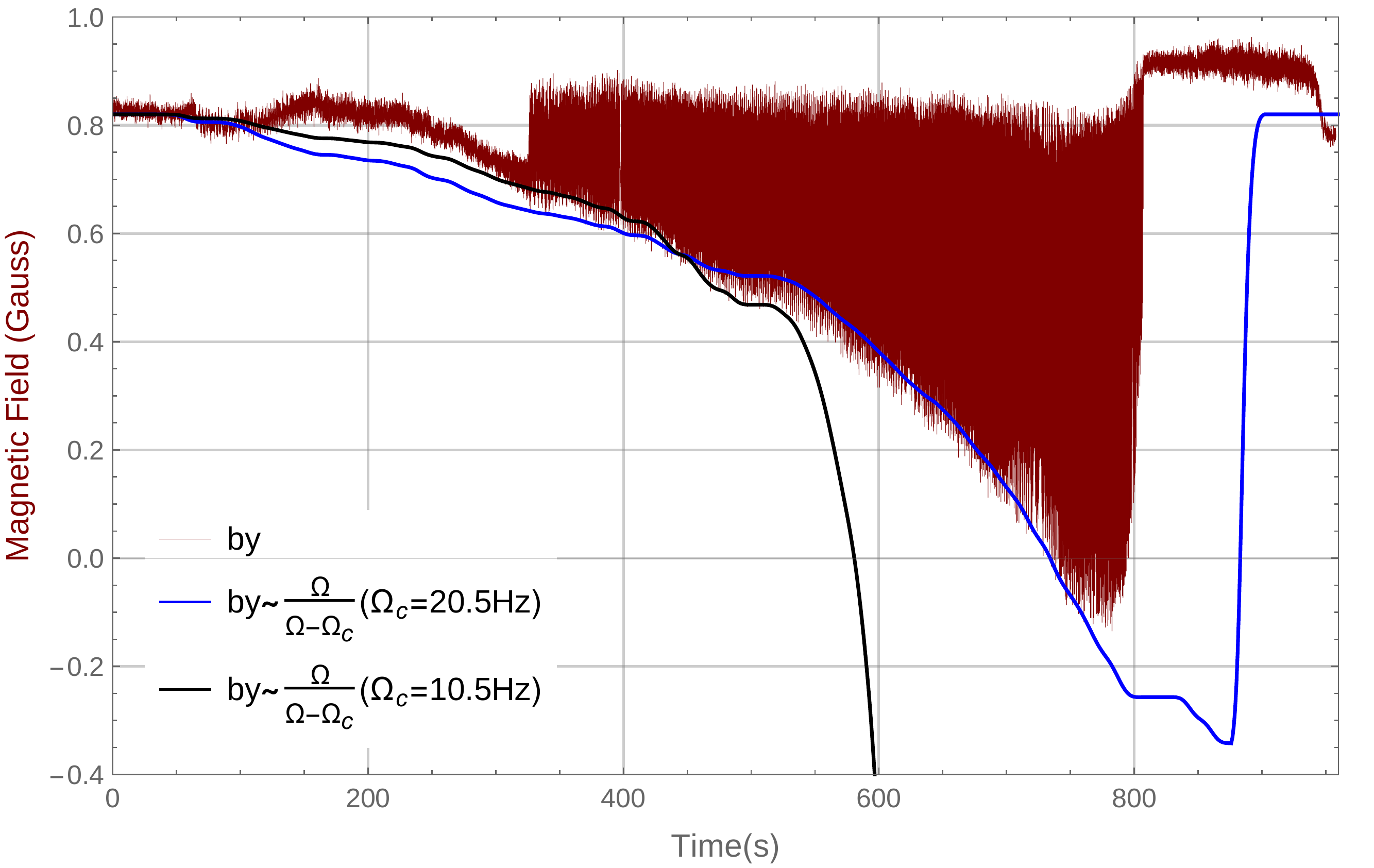}\includegraphics[bb=0bp 0bp 760bp 480bp,clip,width=0.5\textwidth]{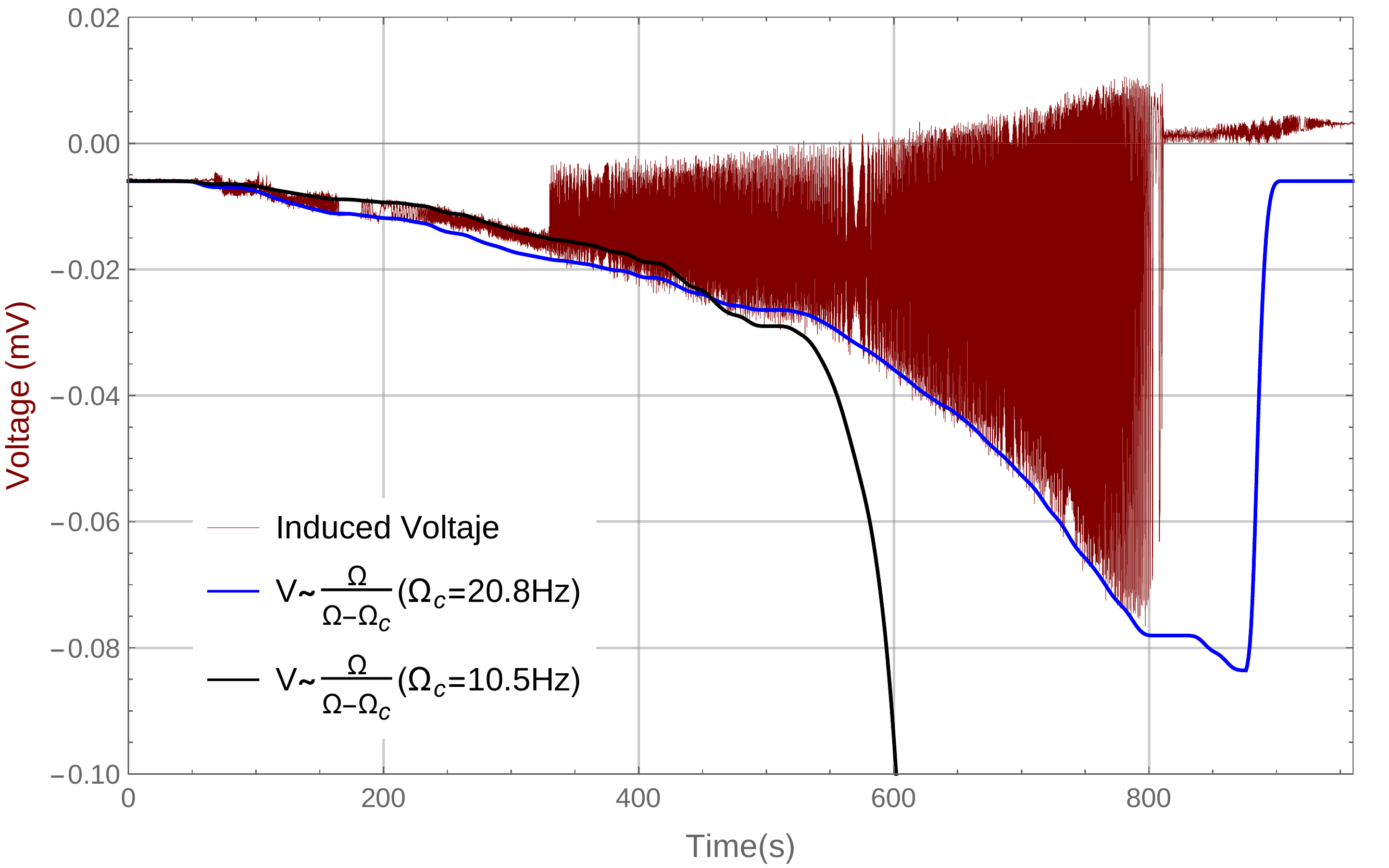}\caption{\label{fig:fit}Comparison of the measured magnetic field and voltage
with the expected variation $\sim\Omega/\left(\Omega_{c}-\Omega\right)$.
The critical rotation rate which best fits the measurements is $\Omega_{c}=\unit[20.5]{Hz}$
( $\Omega_{c}=\unit[20.8]{Hz}$). The black curve corresponds to the
expected theoretical variation function with $\Omega_{c}=\unit[10.5]{Hz}.$}
\end{figure}

On the other hand, combining the equations above we can write the
induced magnetic flux density as 
\[
\phi_{d}=\phi_{0}\Omega/\left(\Omega_{c}-\Omega\right)
\]
 where $\Omega_{c}$ is a critical rotation rate at which the dynamo
becomes self-sustained. A similar expressions hold also for the induced
current and voltage. Using this expression to fit the measurements
in Fig. \ref{fig:fit}, we find $\Omega_{c}\approx\unit[21]{Hz}$
, which is consistent with the results based on the actual contact
resistance shown in Table \ref{tab:omegc}.

\section{Conclusions}

We report the first test results of the homopolar disc dynamo device
proposed by \textit{Priede et al.} \cite{Priede=000026Avalos:PhysLetA:377}
and constructed at CICATA-Querétaro in Mexico. Induced magnetic field
and the voltage drop on the coil were measured for rotation rates
up to $\unit[14]{Hz,}$ at which the liquid metal started to leak
from the outer sliding contact. Although the rotation rate was significantly
above the theoretically predicted dynamo threshold of $\unit[10.4]{Hz,}$
self-excitation of the magnetic field was not achieved. Instead of
the steady magnetic field predicted by the theory we detected a strongly
fluctuating magnetic field with a strength comparable to that of Earth's
magnetic field which was accompanied by similar voltage fluctuations
in the coil. These fluctuations seem to be caused by the intermittent
electrical contact through the liquid metal. The experimental results
suggest that the dynamo with the actual electrical resistance of liquid
metal contacts could be excited at the rotation rate of around $\unit[21]{Hz}$
provided that the leakage of liquid metal is prevented. 

\section*{Acknowledgments}

This work is supported by the National Council for Science and Technology
of Mexico (CONACYT) through the project CB-168850 and by the National
Polytechnic Institute of Mexico (IPN) through the project SIP-20171098.
The first author thanks to M. Valencia, A. Pérez and G. García for
their technical support during the experimental run.

\global\long\def\noopsort{}

\lastpageno 
\end{document}